\documentclass[showpacs,fleqn,nobibnotes]{revtex4}

\usepackage{amsmath}
\usepackage{graphicx}
\graphicspath{{gp/}{t_dist/}{Y_dist/}{ratio/}}
\usepackage[utf8]{inputenc}

\def\lsim{\raise0.3ex\hbox{$<$\kern-0.75em\raise-1.1ex\hbox{$\sim$}}}
\def\gsim{\raise0.3ex\hbox{$>$\kern-0.75em\raise-1.1ex\hbox{$\sim$}}}

\def\pom{{I\!\!P}}
\def\reg{{I\!\!R}}

\newcommand{\rr}{\mbox{\boldmath $r$}}

\newcommand{\rb}{\mbox{\boldmath $b$}}

\newcommand{\rd}{\mbox{\boldmath $\Delta$}}

\newcommand{\N}{\mathcal{N}}

\begin{document}

\title{Probing Saturation Physics in the Real Compton Scattering at Ultraperipheral $pPb$ Collisions}
\author{V. P. Gon\c{c}alves $^{1}$, F. S. Navarra $^{2}$  and D. Spiering $^{2}$}

\affiliation{$^{1}$ Instituto de F\'{\i}sica e Matem\'atica,  Universidade
Federal de Pelotas, 
Caixa Postal 354, CEP 96010-900, Pelotas, RS, Brazil}
\affiliation{$^{2}$ Instituto de F\'{\i}sica, Universidade de S\~{a}o Paulo, CEP 05315-970 S\~{a}o Paulo, SP, Brazil.}

\begin{abstract}
The Real Compton Scattering in ultraperipheral $pPb$ collisions at RHIC and LHC energies is investigated and predictions for the squared transverse momentum ($t$) and rapidity ($Y$) distributions are presented. The scattering amplitude is assumed to be given by the sum of the Reggeon and Pomeron contributions and the Pomeron one is described by the Color Dipole formalism taking into account the non - linear (saturation) effects in the QCD dynamics. We demonstrate that  the behaviour of the cross sections at large -- $t$ and/or $Y$  is dominated by the Pomeron contribution and is strongly affected by the non -- linear effects present in the QCD dynamics. These results indicate that a future experimental analysis of this process can be useful to probe the QCD dynamics at high energies.
\end{abstract}

\maketitle

\section{Introduction}
One important open question in the Standard Model of Particle Physics is the understanding of the strong interaction theory --  Quantum Chromodynamics (QCD) -- in the 
regime of high energies \cite{hdqcd}. In this regime,  hadrons are characterized by a large number of gluons and sea quarks, which form a system with a high partonic 
density. In particular, it is expected that non -- linear interactions, associated to e.g. gluonic recombination $gg \rightarrow g$,  becomes important and should not 
be disregarded in the description of the partonic evolution \cite{glr}. As a consequence, the transition between the linear and non -- linear regimes of  the  QCD dynamics is 
expected to occur with the growth of the energy and/or parton density. This transition is described by the saturation scale $Q_s$, which defines the onset of the 
non -- linear effects. During the last decades, several approaches have been developed to study QCD dynamics at high energies (For recent reviews see, e.g. 
Ref. \cite{hdqcd}). The  formalism of   the  Color Glass Condensate (CGC) \cite{cgc}  is  the current state - of - the - art  of the theory of partons in the high density regime.   
This formalism provides a unified description of inclusive and exclusive observables in $ep/pp/pA/AA$ collisions, which satisfactorily describes the existing experimental data. 
However, as these experimental data can also  be described by alternative approaches, it is important to search for observables and/or processes that can discriminate 
between these distinct descriptions of  QCD at high energies. One of the most promising processes is  exclusive vector meson photoproduction in hadronic collisions, proposed 
originally in Ref. \cite{vicbert} and significantly developed over the last years (See, e.g. Refs. 
\cite{vicmag,vicmag_ups,frankfurt_ups,Schafer_ups,bruno,bruno2,magno,Martin,vicnavdiego,brunorun2} 
). As the amplitude for this process is driven by the gluon content of the target, its cross section  is strongly sensitive to the underlying QCD dynamics. The study of 
exclusive vector meson photoproduction in $pp/pA/AA$ collisions,  as well as other photon -- induced processes \cite{upc}, became a reality in the last years 
\cite{cdf,star,phenix,alice,alice2,lhcb,lhcb2,lhcb3,lhcbconf} and  data from the Run 2 of the LHC are expected to be released soon (For a recent review see 
\cite{review_forward}). 

\begin{figure}[t]
\begin{center}
\scalebox{0.68}{\includegraphics{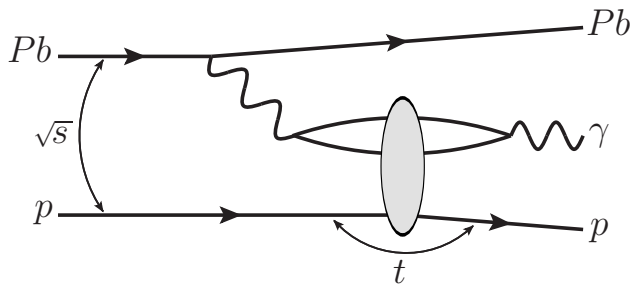}}
\caption{Real Compton scattering in ultraperipheral $pPb$ collisions.}
\label{Fig:dia}
\end{center}
\end{figure}

The studies performed in Refs. \cite{bruno,bruno2,brunorun2}  have demonstrated that the experimental LHC Run 1 data and the preliminary Run 2 data on exclusive vector meson 
photoproduction can be sucessfully described with the color dipole formalism if  non - linear effects in the QCD dynamics are taken into account. One of the  main advantages  
of this approach is that the basic ingredients can be constrained by the very precise HERA data and hence the predictions for photon -- induced interactions at the LHC are 
parameter free. Another advantage is that it can be easily extended to describe other final reactions as  diffractive heavy quark production \cite{vicmaghqd}, exclusive $Z^0$ 
photoproduction \cite{vicmagz0,motwatt} and the time -- like Compton scattering \cite{tlcompton1,tlcompton3,tlcompton2}, providing a unified description of all these exclusive 
processes. In this paper we extend the color dipole formalism to treat the Real Compton Scattering in ultraperipheral $pPb$ collisions (See Fig. \ref{Fig:dia})  and estimate 
the corresponding  rapidity and squared  momentum transfer ($t$) distributions at RHIC and LHC energies. Our analysis is motivated by the study performed in Ref. \cite{kope_rcs} 
using a distinct approach. As we will demonstrate in what follows, our results indicate that the total cross section is dominated by the  Reggeon (non -- perturbative) contribution.  
On the other hand,  at large -- $t$, the cross section is dominated by the Pomeron contribution, which is described by the dipole approach taking into account  non -- linear 
effects. Moreover, we show that the Pomeron contribution also dominates exclusive  photon production at large rapidities $Y$.  Finally, we demonstrate that the behavior of the 
differential cross section at large values of  $t$ and $Y$ is strongly affected by the non -- linear effects present in the QCD dynamics.  Our results indicate that a future 
experimental analysis can be useful to probe the non - linear (saturation) physics.

This paper is organized as follows. In the next Section we present a brief review of the formalism used to describe the Real Compton Scattering in hadronic collisions. In 
Section \ref{res} we will present our predictions for the rapidity and  $t$ -- distributions and in Section \ref{conc} we will summarize our main conclusions.

\section{Formalism}

In a hadronic collision at high energies, both incident charged particles  give rise to strong electromagnetic fields \cite{upc}. Consequently, when they collide at large 
impact parameters, the interaction between them will be dominated by photon --  induced processes. In particular,  the photon stemming from the electromagnetic field of one 
of the two colliding hadrons can interact with a photon coming from the other hadron (photon - photon process) or it can interact directly with the other hadron 
(photon - hadron process). In what follows we will focus on the real Compton process initiated by exclusive photon -- hadron interactions. In this process, both  incident 
hadrons will remain intact  and two rapidity gaps will appear in the final state, in both sides of the produced photon (See Fig. \ref{Fig:dia}). Moreover, as the photon flux 
of  a nucleus is enhanced by a factor $Z^2$ in comparison to the proton one,  in a $pPb$ collision, the process will be dominated by photons coming from the nucleus.  
The  differential cross section will be given by
\begin{widetext}
\begin{eqnarray}
  \frac{d\sigma \,\left[\text{Pb} + p \rightarrow \text{Pb} \otimes \gamma \otimes p\right]}{dY\,dt}  = 
   \left[\omega \frac{dN}{d\omega}\bigg|_{\text{Pb}}\,\frac{d\sigma}{dt}(\gamma p \rightarrow \gamma \otimes p)\right]\,\,, 
\label{dsigdy}
\end{eqnarray}
\end{widetext}
where  $\omega$ is the energy of the incident photon in the collider frame, 
 $Y$ is the rapidity of the photon in the final state and $\frac{dN}{d\omega}$ denotes the  equivalent photon 
spectrum  of the relativistic incident nucleus. In our calculations, we will assume that the nucleus is moving from the left to the right, which implies that  positive rapidities correspond to the nucleus fragmentation region. As in Refs.  \cite{bruno,brunorun2} we will assume that the photon flux associated to the   nucleus  can be described by   
the relativistic point -- like charge  model \cite{upc}. Moreover, $d\sigma/dt$ is the differential cross section of the $\gamma p \rightarrow \gamma \otimes p$ process, 
with the symbol $\otimes$ representing the presence of a rapidity gap in the final state.

The differential cross section for the Real Compton process ($\gamma + p \rightarrow \gamma + p$)  can be expressed as follows
\begin{eqnarray}
\frac{d\sigma}{dt}
& = & \frac{1}{16\pi}  |{\cal{A}}^{\gamma p \rightarrow \gamma p }(W^2,  \Delta)|^2\,\,,
\label{dsigdt}
\end{eqnarray}
where $\Delta = - \sqrt{t}$ is the momentum 
transfer and $W^2$ is the squared center -- of -- mass energy. In what follows we will assume that the scattering amplitude is given by
\begin{eqnarray}
{\cal{A}}^{\gamma p \rightarrow \gamma p }(W^2,  \Delta) =  {\cal{A}}^{\reg}(W^2,\,\Delta) + {\cal{A}}^{\pom}(W^2,\,\Delta) \,\,,
\end{eqnarray}
where $\reg$ and $\pom$ denote the Reggeon and Pomeron contributions, which  determine the energy behavior of the cross section at low and high energies, respectively. 
In order to get the scattering amplitude ${\cal{A}}$, we will use the fact that its imaginary part  determines, throught the optical theorem, the total cross section of the 
process $\gamma + p \rightarrow X$.
Following previous studies \cite{timeanu,vicgratierihq,kope_rcs}, we will assume that $\sigma_{\gamma p \rightarrow X}$ can be expressed by
\begin{eqnarray}
  \sigma_{\gamma p}(W^2) =  \sigma_{\gamma p}^\reg(W^2) + \sigma_{\gamma p}^\pom(W^2) \,\,.
\end{eqnarray}
As in Ref. \cite{timeanu} we will assume that ${\cal{A}}^{\reg}$ is given by 
\begin{eqnarray}
  {\cal A}^{\gamma p \rightarrow \gamma p }_\reg(W^2,\,\Delta)  =  16\pi\sigma_\reg(W^2)\,e^{B_\reg t}
\label{amp_reg}
\end{eqnarray}
with $ \sigma_{\gamma p}^\reg(W^2) = 0.135 \cdot (W^2)^{-3}$ mb \cite{timeanu}, 
$  B_\reg(s) = B_1 + 2\alpha_\reg'(0)\ln(W^2/\mu_0^2)$, $B_1=6\text{ GeV}^{-2}$, $\alpha_\reg'(0)=0.9\text{ GeV}^{-2}$ and $\mu_0^2=1\text{ GeV}^{2}$ \cite{kope_rcs}.
Moreover, we will assume that the Pomeron contribution can be described by the dipole model \cite{nik}. Consequently, the  associated scattering amplitude for the 
$\gamma + p \rightarrow \gamma + p$ process and the total $\gamma + p \rightarrow X$  cross section can be expressed, respectively, by
\begin{widetext}
\begin{eqnarray}
  {\cal A}^{\gamma p \rightarrow \gamma p }_{\pom}(W^2,\,\Delta)  =  i \sum_f\int dz \, d^2\rr \, d^2\rb \, 
  e^{-i[\rb-(1/2-z)\rr].\rd} \,\left[ \Psi^*_\gamma\Psi_\gamma (z,\rr) \right]^f_T 2\N(x,\,\rr,\,\rb)
\label{amp_dip}
\end{eqnarray}
and 
\begin{eqnarray}
  \sigma_{\gamma p}^\pom(W^2) = \sum_f \int dz\,d^2\rr \,d^2\rb \left[ \Psi^*_\gamma\Psi_\gamma (z,\rr) \right]^f_T 2\N(x,\,\rr,\,\rb)\,\,,
\end{eqnarray}
\end{widetext}
where    $\rr$ and $z$ are the dipole transverse radius and the momentum fraction of the photon carried by the quark (the antiquark carries then $1-z$), respectively. Moreover, the Bjorken -- $x$ variable will be given by $x = 4m_f^2/W^2$, with   $m_f$ being the  mass of the quark of  flavor $f$.
Furthermore,  $\rb$ is the impact parameter of the dipole relative to the proton and  $ {\cal N} (x, \rr, \rb)$ is the 
forward dipole -- proton scattering amplitude (for a dipole at  impact parameter $\rb$) which encodes all the information about the hadronic scattering, 
and thus about the non-linear and quantum effects in the hadron wave function (see below).
Finally,   $[\Psi^{*}_\gamma\Psi_\gamma]^f_T$ denotes the photon wave function overlap for an initial photon with tranverse polarization, which can be calculated using Quantum 
Electrodynamics and  which is given by
\begin{widetext}
\begin{eqnarray}
  \left[ \Psi^*_\gamma\Psi_\gamma (z,\rr) \right]^f_T = \frac{3\alpha_{em}e_f^2}{2\pi^2}
  \left\{\left[z^2 + (1-z^2)\right]m_f^2K_1^2(m_fr) + m_f^2K_0^2(m_fr)\right\}\,\,.
  \label{wave}
\end{eqnarray}
\end{widetext}
As in Refs. \cite{bruno,bruno2,brunorun2}, we will use  the impact parameter Color Glass Condensate (bCGC) model \cite{KT,KMW} for the dipole -- proton 
scattering amplitude ${\cal{N}}$, which is given by \cite{KMW} 
\begin{widetext}
\begin{eqnarray}
\mathcal{N}(x,\rr,\rb) =   
\left\{ \begin{array}{ll} 
{\mathcal N}_0\, \left(\frac{ r \, Q_s(b)}{2}\right)^{2\left(\gamma_s + 
\frac{\ln (2/r \, Q_s(b))}{\kappa \,\lambda \,Y}\right)}  & \mbox{$r Q_s(b) \le 2$} \\
 1 - e^{-A\,\ln^2\,(B \, r \, Q_s(b))}   & \mbox{$r Q_s(b)  > 2$} 
\end{array} \right.
\label{eq:bcgc}
\end{eqnarray}
\end{widetext} 
with  $\kappa = \chi''(\gamma_s)/\chi'(\gamma_s)$, where $\chi$ is the 
LO BFKL characteristic function.  The coefficients $A$ and $B$  
are determined uniquely from the condition that $\mathcal{N}(x,\rr,\rb)$, and its derivative 
with respect to $r\,Q_s(b)$, are continuous at $r\,Q_s(b)=2$. The behavior of $\mathcal{N}$ in the kinematical range  $r Q_s(b) \le 2$ ($r Q_s(b) >  2$) describes 
the QCD dynamics in the linear (non -- linear) regime. The impact parameter dependence of the  proton saturation scale $Q_s(b)$  is given by:
\begin{equation} 
  Q_s(b)\equiv Q_s(x,b)=\left(\frac{x_0}{x}\right)^{\frac{\lambda}{2}}\;
\left[\exp\left(-\frac{{b}^2}{2B_{\rm CGC}}\right)\right]^{\frac{1}{2\gamma_s}},
\label{newqs}
\end{equation}
with the parameter $B_{\rm CGC}$  being obtained by a fit of the $t$-dependence of 
exclusive $J/\psi$ photoproduction. The  factors $\mathcal{N}_0$ and  $\gamma_s$  were  
taken  to be free. In what follows we consider the set of parameters obtained in 
Ref. \cite{amir} by fitting the recent HERA data on the reduced $ep$ cross sections:
 $\gamma_s = 0.6599$, $\kappa = 9.9$, $B_{CGC} = 5.5$ GeV$^{-2}$, $\mathcal{N}_0 = 0.3358$, $x_0 = 0.00105$ and $\lambda = 0.2063$.
As demonstrated in Ref. \cite{amir},  these models allow us to successfully describe  the high precision combined 
HERA data on inclusive and exclusive processes. Finally, in order to investigate the impact of the non -- linear (saturation) effects on the real Compton scattering, 
we  will also estimate the scattering amplitude assuming that  $\mathcal{N}(x,\rr,\rb)$ is given by the linear part of the bCGC model, which is 
\begin{eqnarray}
\mathcal{N}(x,\rr,\rb) =  
{\mathcal N}_0\, \left(\frac{ r \, Q_s(b)}{2}\right)^{2\left(\gamma_s + 
\frac{\ln (2/r \, Q_s(b)}{\kappa \,\lambda \,Y}\right)}\,,
\label{eq:bcgclin}
\end{eqnarray}
with the same parameters used before in Eq. (\ref{eq:bcgc}). The Pomeron contributions to the scattering amplitude 
${\cal A}^{\gamma p \rightarrow \gamma p }$ and to $\sigma_{\gamma p \rightarrow X}$ are fully determined (in the dipole approach) by the overlap function and 
by $\mathcal{N}$, which has its  parameters constrained by the inclusive and exclusive $ep$ HERA data. Consequently, the predictions for ultraperipheral collisions are 
parameter free.

\begin{figure}[t]
\begin{center}
\scalebox{0.35}{\includegraphics{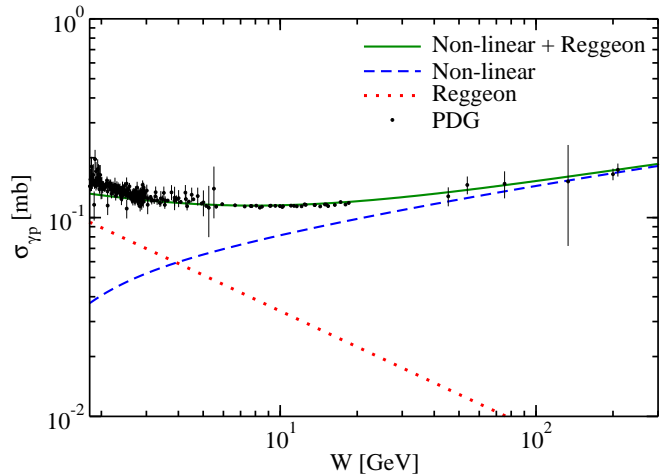}}
\caption{The total $\gamma p$ cross section as a function of the center-of-mass energy. Data from Ref. \cite{PDG_2012}.}
\label{Fig:gp_total}
\end{center}
\end{figure}

\section{Results}
\label{res}
Our goal in this Section is to present our results for the real Compton scattering in ultraperipheral $pPb$ collisions. However, before discussing them, it is important to 
demonstrate that with our approach we can describe the experimental data on the total $\gamma p$ cross section, which is  related to the scattering amplitude of the real 
Compton scattering through the optical theorem.  In Fig. \ref{Fig:gp_total} we compare our predictions with the existing data for center - of - mass energies $W$ larger 
than 2 GeV. The Reggeon and Pomeron contributions are presented separately. The latter will be denoted ``Non - linear'' hereafter and will be calculated with the full bCGC model. 
As in Ref. \cite{vicgratierihq}, we will assume that $m_u = m_d = m_s = 0.2 $ GeV and $m_c = 1.5$ GeV.  As expected, the Reggeon contribution decreases with the energy, and 
the high energy behavior of the cross section is determined by the Pomeron term. Moreover,  the bCGC model is able to describe the high energy H1 and ZEUS data on the total 
photoproduction cross section. This result justifies the application of the color dipole model to the real Compton scattering at large energies.

\begin{figure}[t]
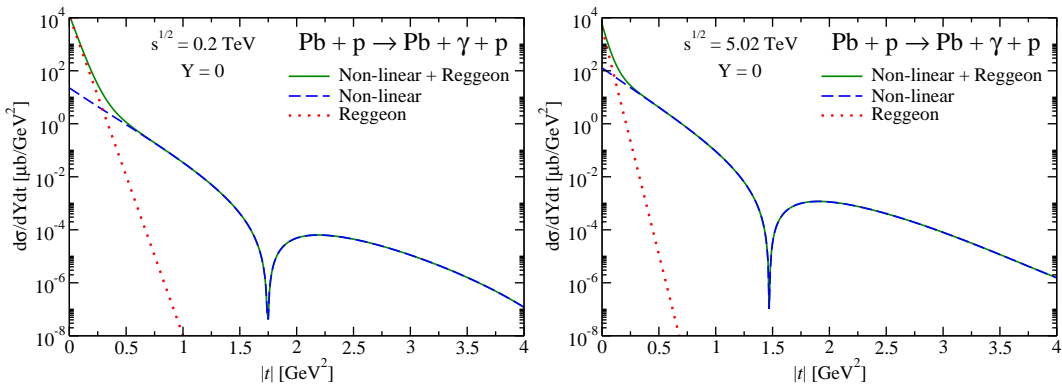

\begin{center}
\scalebox{0.28}{\includegraphics{dsdtdY_pA_bCGC_t_dist_0_2_b.eps}}
\scalebox{0.28}{\includegraphics{dsdtdY_pA_bCGC_t_dist_5_02_b.eps}}
\caption{Transverse momentum distributions in the exclusive real photon production at central rapidities ($Y=0$) in ultraperipheral $pPb$ collisions at 
RHIC (left panel)  and LHC (right panel) energies.}
\label{Fig:t_dist}
\end{center}
\end{figure}

\begin{figure}[t]
\begin{center}
\scalebox{0.35}{\includegraphics{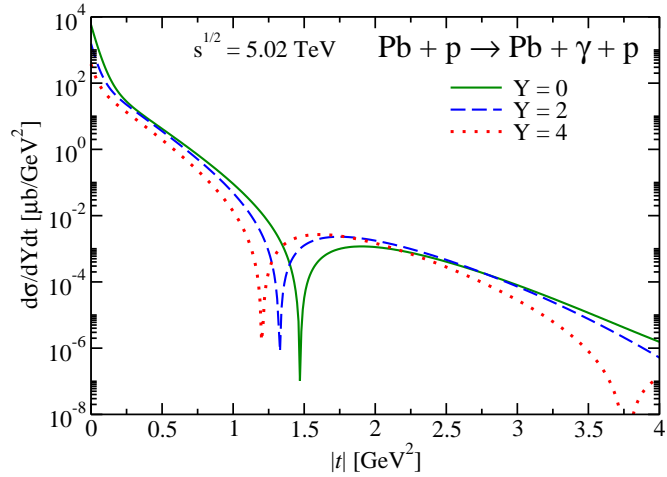}}
\caption{Rapidity dependence of the transverse momentum distributions in ultraperipheral $pPb$ collisions at LHC  energy.}
\label{Fig:t_dist_rap}
\end{center}
\end{figure}

Let us first estimate the transverse momentum distributions in exclusive real photon production at central rapidities ($Y=0$) in ultraperipheral $pPb$ collisions at 
RHIC ($\sqrt{s} = 0.2$ TeV)  and LHC ($\sqrt{s} = 5.02$ TeV) energies.  Our results for the Reggeon, Non -- linear and full (Reggeon + Non -- linear) predictions are 
presented in Fig. \ref{Fig:t_dist}. They clearly indicate that the total cross section, which is determined by the behavior at $|t| = 0$,  is dominated by the  
soft interactions described by the Reggeon contribution. On the other hand, at large - $|t|$, the behavior of the distribution is determined by the Pomeron (Non -- linear) 
contribution. At higher energies the Pomeron dominance becomes more and more pronounced.  Moreover, we predict a dip in the distribution. The position of the dip   
moves to  smaller values of $|t|$ as the  energy increases, similarly to what we predicted for  the vector meson photoproduction in ultraperipheral collisions 
\cite{vicnavdiego}. Moreover, the results presented in Fig. \ref{Fig:t_dist_rap} indicate that the position of the dip is shifted at smaller values of $|t|$ when the rapidity is increased.  These results indicate that the experimental analysis of this process at large - $|t|$ can be useful to investigate the properties of  QCD at high 
energies.

\begin{figure}[t]
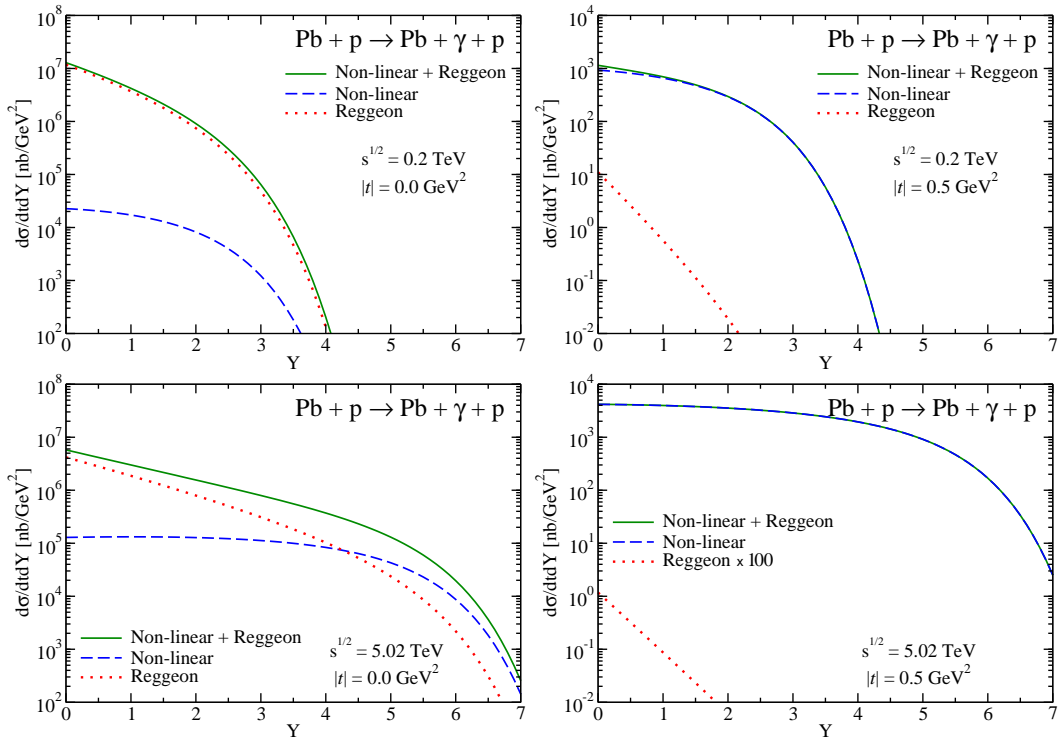

\begin{center}
\scalebox{0.28}{\includegraphics{dsdtdY_pA_bCGC_Y_dist_s_0_2_t_0_0.eps}}
\scalebox{0.28}{\includegraphics{dsdtdY_pA_bCGC_Y_dist_s_0_2_t_0_5.eps}}\\
\scalebox{0.28}{\includegraphics{dsdtdY_pA_bCGC_Y_dist_s_5_02_t_0_0.eps}}
\scalebox{0.28}{\includegraphics{dsdtdY_pA_bCGC_Y_dist_s_5_02_t_0_5.eps}}
\caption{Rapidity distribution in exclusive real photon production in ultraperipheral $pPb$ collisions at RHIC (upper panels)  and LHC (lower panels) energies. 
Results for $|t| = 0$ (left panels) and 0.5 GeV$^2$ (right panels).}
\label{Fig:Y_dist_rhic}
\end{center}
\end{figure}

In Fig. \ref{Fig:Y_dist_rhic} we present our predictions for the rapidity distributions to be measured in  ultraperipheral $pPb$ collisions at RHIC (upper panels)  
and LHC (lower panels) energies. For comparison, we present our results for $|t| = 0$ (left panels) and 0.5 GeV$^2$ (right panels).  Our results indicate that at 
RHIC energies and forward scattering ($|t| = 0$), the process will be dominated by the Reggeon contribution in the full kinematical range. In contrast, at 
$|t|$ = 0.5 GeV$^2$, the behavior of the rapidity distribution is determined by the QCD dynamics at high energies. At LHC energies and $|t|$ = 0,  that Non -- linear 
contribution is similar to the Reggeon one in the kinematical range probed by the LHCb detector and  dominates at forward rapidities. For $|t|$ = 0.5 GeV$^2$, the 
Reggeon contribution is strongly suppressed, with the distribution being a direct probe of  QCD dynamics.

\begin{figure}[t]
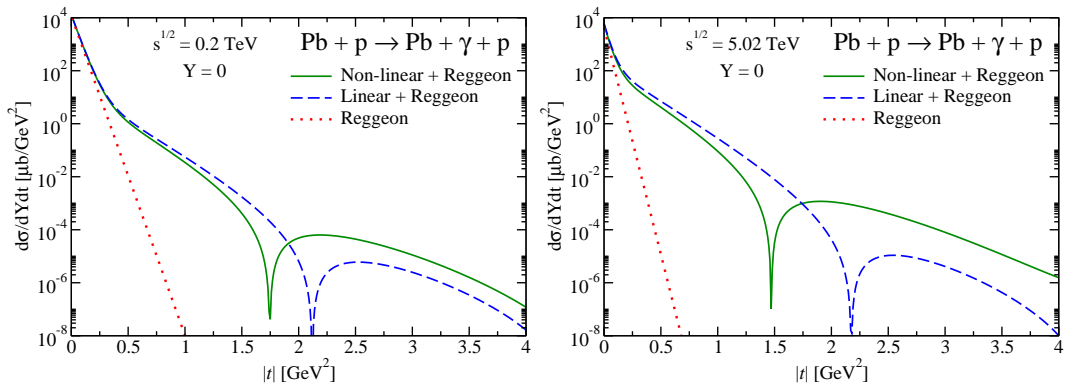

\begin{center}
\scalebox{0.28}{\includegraphics{dsdtdY_pA_bCGC_t_dist_0_2.eps}}
\scalebox{0.28}{\includegraphics{dsdtdY_pA_bCGC_t_dist_5_02.eps}}
\caption{Comparison between the linear and non -- linear predictions for the transverse momentum distributions at $Y$ = 0 in ultraperipheral $pPb$ collisions 
at RHIC (left panel)  and LHC (right panel) energies.}
\label{Fig:t_dist_linear}
\end{center}
\end{figure}

\begin{figure}[t]
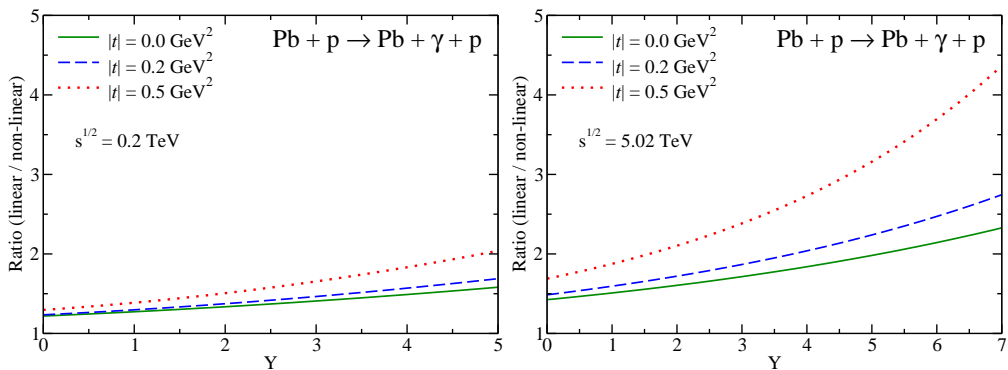

\begin{center}
\scalebox{0.28}{\includegraphics{dsdtdY_pA_bCGC_Y_dist_ratio_s_0_2.eps}}
\scalebox{0.28}{\includegraphics{dsdtdY_pA_bCGC_Y_dist_ratio_s_5_02.eps}}
\caption{Rapidity dependence of the ratio between the linear and non-linear predictions for the rapidity distributions for RHIC (left panel) and LHC (right panel) 
energies assuming different values of $|t|$.}
\label{Fig:ratio}
\end{center}
\end{figure}

Let us  now investigate the impact of  non -- linear effects on real Compton scattering in ultraperipheral $pPb$ collisions. In order to do that, we will compare our previous results, derived using Eq. (\ref{eq:bcgc}) to estimate the Pomeron contribution, with those obtained using Eq. (\ref{eq:bcgclin}) as 
input in the calculations. In Fig. \ref{Fig:t_dist_linear} we present our predictions for $Y = 0$. The dominance of the Reggeon contribution at $|t| \approx 0$  
is not modified by the properties of QCD dynamics at high energies. In contrast, if the non -- linear effects are disregarded, the position of the dip is modified, 
occuring at larger values of $|t|$. A similar behavior  was predicted in Ref. \cite{vicnavdiego} for  exclusive vector meson photoproduction. 
Therefore, the measurement of the $t$ -- distribution can be considered an important probe of the  QCD dynamics.
Additionally, the results presented in Fig. \ref{Fig:ratio} for the ratio between the linear and non -- linear predictions for the rapidity distributions indicate 
that the impact of non -- linear effects is larger  at forward rapidities, increasing with the value of $|t|$ considered. Finally, it is important to emphasize that 
we have checked that similar conclusions are valid for the Run 2 energy of the LHC ($\sqrt{s} = 8$ TeV).

\begin{figure}[t]
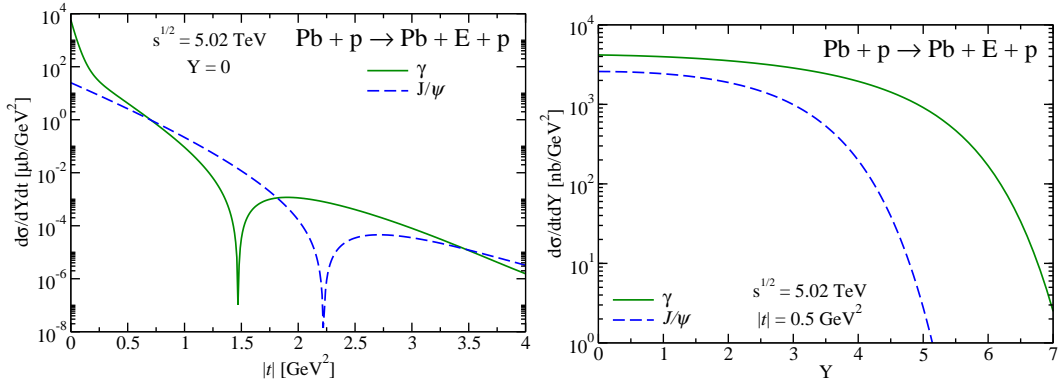

\begin{center}
\scalebox{0.28}{\includegraphics{dsdtdY_pA_bCGC_t_dist_5020_meson2.eps}}
\scalebox{0.28}{\includegraphics{dsdtdY_pA_bCGC_Y_dist_s=5020_t=05_meson2.eps}}
\caption{Comparison between the predictions for the Real Compton scattering and exclusive $J/\Psi$ photoproduction in  ultraperipheral $pPb$ collisions 
at LHC energy.}
\label{Fig:t_dist_comps}
\end{center}
\end{figure}

Finally, in Fig. \ref{Fig:t_dist_comps} we compare our predictions for the Real Compton Scattering with those for  the exclusive $J/\Psi$ photoproduction in  
ultraperipheral $pPb$ collisions 
at LHC energy. The results for the $J/\Psi$ production are complementary those presented in Ref. \cite{vicnavdiego} for $pp$ and $PbPb$ collisions. We have that the total RCS cross section is larger than the $J/\Psi$ one, the dip in the $|t|$ -- distribution  occurs at smaller values of $|t|$ and the rapidity distribution is broader. 
These results indicate that the study of the  Real Compton Scattering is is principle feasible at the LHC. However, the experimental separation of the produced photons still is a challenge, which deserves more detailed studies. 

\section{Conclusions}
\label{conc}
In this paper we have investigated the real Compton scattering in ultraperipheral $pPb$ collisions. Our study is strongly motivated by the fact that exclusive processes 
in photon -- induced interations at hadronic collisions  are important probes of the QCD dynamics at high energies. Assuming that the scattering amplitude can be expressed 
by the sum of the Reggeon and Pomeron contributions and that the Pomeron one can be described by the color dipole formalism, we demonstrated that with this approach we can 
reproduce the experimental data on  $\sigma_{\gamma p \rightarrow X}$. More importantly,  with this approach we obtain parameter free predictions for the real Compton 
scattering. Predictions for the transverse momentum and rapidity distributions have been presented taking into account non -- linear (saturation) effects  
in the QCD dynamics. We demonstrated that  the behavior of the cross sections at large -- $t$ and/or $Y$  is dominated by the Pomeron contribution and it is strongly 
affected by the non -- linear effects inherent to the QCD dynamics.  These results are robust  predictions of  saturation physics. Our results indicate that 
a future experimental analysis of real Compton scattering in ultraperipheral $pPb$ collisions can be useful to probe the QCD dynamics at high energies.

\begin{acknowledgements}
 This work was  partially financed by the Brazilian funding agencies CNPq, CAPES, FAPESP and FAPERGS.
\end{acknowledgements}

\end{document}